\documentclass[letterpaper,10pt]{article}
\usepackage{osameet2}

\usepackage{tikz}
\usepackage{pgfplots}
\pgfplotsset{compat=newest} 
\pgfplotsset{plot coordinates/math parser=false}
\usepackage{graphicx}
\usepackage{color}
\usepackage{amsmath,amsthm}
\usepackage{amssymb, bm, dsfont}
\usepackage{float}
\usepackage{stmaryrd,amsmath,amsfonts,rotating}
\usepackage{booktabs}
\usepackage{wrapfig}

\usepackage{amsmath,amssymb}

\theoremstyle{definition}

\begin{document}

\title{An Efficient Nonlinear Fourier Transform Algorithm for Detection of Eigenvalues from Continuous Spectrum }

\author{Vahid Aref, Son T. Le, Henning Buelow}
\address{Nokia Bell Labs, Stuttgart, Germany}
\email{vahid.aref@nokia-bell-labs.com}

\begin{abstract}
We present an efficient, fast and robust Nonlinear Fourier Transform (NFT) algorithm to 
detect eigenvalues of the discrete spectrum. It outperforms other known NFT algorithms 
as it detects the eigenvalues from the continuous spectrum, the numerically more robust part of 
the nonlinear spectrum.
\end{abstract}

\section{Introduction}

Nonlinear Frequency Division Multiplexing (NFDM) has been proposed as a viable technique 
to exploit the Kerr nonlinearity in data modulation over nonlinear optical fiber~\cite{yousefi2014information}.
In NFDM systems, the data is modulated in a so-called nonlinear Fourier spectrum~\cite{le2017nature}.
The spectrum has two parts:
the continuous spectrum containing the real valued frequencies 
and the discrete spectrum containing a set of isolated complex valued frequencies, called eigenvalues. 
The discrete spectrum represents the solitonic components of the signal. 
The main advantage of NFDM is that 
the nontrivial transformation of a pulse along an ``ideal'' nonlinear optical fiber can be
characterized by simple linear transfer functions in the nonlinear Fourier spectrum. Various NFDM systems
have been shown in the last years in different scenarios, see
\cite{turitsyn2017nonlinear} and references therein.


To retrieve transmitted data, the received signal should be mapped to its nonlinear Fourier spectrum. 
There are plenty of Nonlinear Fourier transform (NFT) algorithms, see \cite{yousefi2014information,turitsyn2017nonlinear}, to numerically 
compute the spectrum. While the algorithms for the continuous spectrum are robust and relatively precise,
the ones for the discrete spectrum suffer from 
severe problems, summerized in Sec.~\ref{sec:nft}, specially
when the signal duration is relatively long.
This is the case when the pulse has several eigenvalues~\cite{Buelow20167eigenvalues} or the pulse has both discrete and continuous spectrum~\cite{le2017nature},\cite{aref2018modulation}. 

In this paper, we present a novel algorithm to retrieve the eigenvalues of the discrete spectrum from the continuous spectrum. The algorithm exploits the relation between these two spectrum to find all discrete eigenvalues simultaneously. We show its excellent performance on the pulses reported before in an NFDM experiment~\cite{aref2018modulation}. 
Each pulse has 4 eigenvalues, modulated 8-PSK independently, as well as a modulated continuous spectrum by a
64x0.5 Gbaud OFDM signal with 32-QAM sub-carriers, resulting 55.3 Gb/s.
As we show in Sec.~\ref{sec:exp}, the
relatively long duration of each pulse makes it
challenging for other known NFT algorithms to retrieve eigenvalues from the received pulses.

\section{Nonlinear Fourier Transform: Preliminaries and Numerical Problems}\label{sec:nft}

The standard Nonlinear Schr{\"o}dinger Equation (NLSE) serves as
the basic model for the pulse propagation $q(t,z)$ along an ideally lossless and noiseless fiber.
The nontrivial pulse propagation can be characterized
by simple transformations in the nonlinear Fourier spectrum, defined by the following so-called Zakharov-Shabat system\footnote{The Zakharov-Shabat problem is usually defined differently, e.g. \cite{ablowitz1981solitons}. This equivalent but simpler form is obtained by change of variables.}~\cite{ablowitz1981solitons}
\begin{equation}\label{eq:ZS}
\frac{\partial}{\partial t}\left(\begin{matrix}v_1(t;\lambda,z)\\ v_2(t;\lambda,z)\end{matrix}\right)=
	\left(\begin{matrix}
	0 & q\left(t,z\right)e^{+2j\lambda t} \\-q^*\left(t,z\right)e^{-2j\lambda t} & 0
	\end{matrix}\right)
	\left(\begin{matrix}v_1(t;\lambda,z)\\ v_2(t;\lambda,z)\end{matrix}\right),\hspace{1cm}
	\lim_{t\to-\infty} \left(\begin{matrix} v_1(t;\lambda,z)\\v_2(t;\lambda,z)\end{matrix}\right)=\left(\begin{matrix}1\\0\end{matrix}\right).
\end{equation}
The nonlinear Fourier coefficients (Jost pair) are then defined as
\begin{equation*}
a\left(\lambda;z\right)=\lim_{t\to+\infty} v_1(t;\lambda,z), b\left(\lambda;z\right)=\lim_{t\to+\infty} v_2(t;\lambda,z).
\end{equation*}
The nonlinear spectrum is usually described by the following two parts: \\$(i)$ 
Continuous spectrum (CS): the spectral amplitude $Q_c(\lambda;z)=b(\lambda;z)/a(\lambda;z)$ for real frequencies $\lambda\in\mathbb{R}$. \\$(ii)$ Discrete spectrum (DS):
$\{\lambda_k,Q_d(\lambda_k;z)\}$ where 
$\lambda_k\in\mathbb{C}^+$ (upper complex plane) such that $a(\lambda_k;z)=0$. The spectral amplitudes are defined as $Q_d(\lambda_k;z)=b(\lambda_k;z)/\frac{\partial a(\lambda;z)}{\partial \lambda}|_{\lambda=\lambda_k}$.

Look at Eq.~\ref{eq:ZS}.
For $\lambda\in\mathbb{R}$, the matrix is skew-Hermitian and the off-diagonal entries are bounded.
These properties allow for designing robust algorithms with quasi-linear complexity~\cite{wahls2015fast}.
The NFT algorithms for DS are reviewed in \cite{yousefi2014information,vasylchenkova2018direct}. 
All the algorithms, based on discretizing \eqref{eq:ZS} or its variants, suffer, more or less, from the following problems:\\
\begin{wrapfigure}{r}{0.46\textwidth}
    \begin{minipage}{1\linewidth}
    \centering
\begingroup%
  \makeatletter%
  \providecommand\color[2][]{%
    \errmessage{(Inkscape) Color is used for the text in Inkscape, but the package 'color.sty' is not loaded}%
    \renewcommand\color[2][]{}%
  }%
  \providecommand\transparent[1]{%
    \errmessage{(Inkscape) Transparency is used (non-zero) for the text in Inkscape, but the package 'transparent.sty' is not loaded}%
    \renewcommand\transparent[1]{}%
  }%
  \providecommand\rotatebox[2]{#2}%
  \ifx\svgwidth\undefined%
    \setlength{\unitlength}{01\textwidth}%
    \ifx\svgscale\undefined%
      \relax%
    \else%
      \setlength{\unitlength}{\unitlength * \real{\svgscale}}%
    \fi%
  \else%
    \setlength{\unitlength}{\svgwidth}%
  \fi%
  \global\let\svgwidth\undefined%
  \global\let\svgscale\undefined%
  \makeatother%
  \begin{picture}(1,0.57210699)%
    \put(0,0){\includegraphics[width=\unitlength]{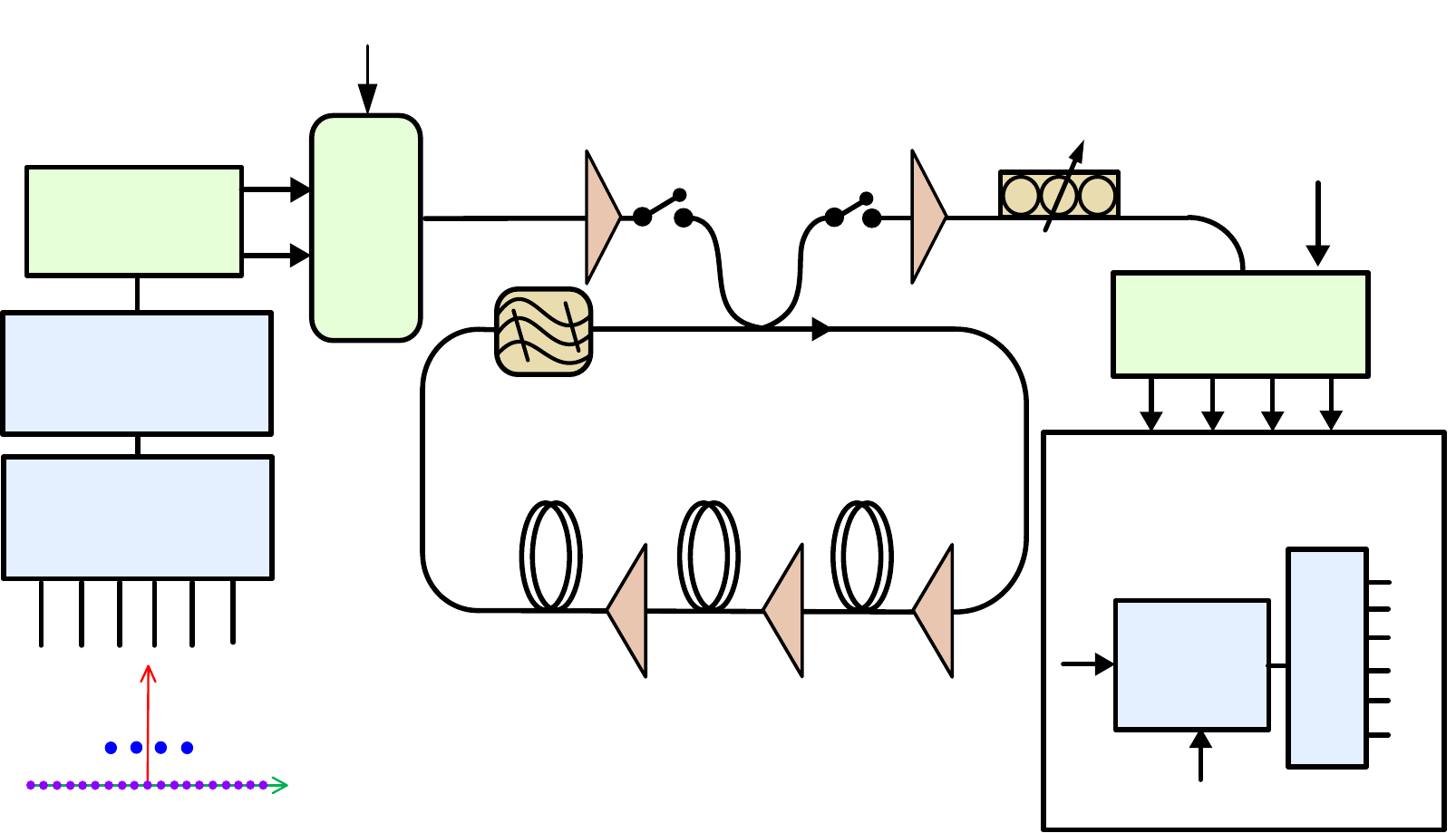}}%
     \scriptsize
    \put(0.70634307,0.48){\color[rgb]{0,0,0}\makebox(0,0)[lb]{\smash{PC}}}%
    \put(0.17944123,0.45){\color[rgb]{0,0,0}\makebox(0,0)[lb]{\smash{I}}}%
    \put(0.17378229,0.41){\color[rgb]{0,0,0}\makebox(0,0)[lb]{\smash{Q}}}%
    \put(0.60,0.07){\color[rgb]{0,0,0}\makebox(0,0)[lb]{\smash{EDFA}}}%
    \put(0.89692283,0.0907171){\color[rgb]{0,0,0}\rotatebox{90}{\makebox(0,0)[lt]{\begin{minipage}{0.03363347\unitlength}\centering NFT\end{minipage}}}}%
    \put(0.70493367,0.26){\color[rgb]{0,0,0}\makebox(0,0)[lt]{\begin{minipage}{0.29217743\unitlength}\centering offline processing\end{minipage}}}%
    \put(0.78,0.14662738){\color[rgb]{0,0,0}\makebox(0,0)[lt]{\begin{minipage}{0.07067136\unitlength}\centering  Normal\\ ization\end{minipage}}}%
    \put(0.69308738,0.1990669){\color[rgb]{0,0,0}\makebox(0,0)[lt]{\begin{minipage}{0.13759147\unitlength}\centering $r(t)$\end{minipage}}}%
    \put(-0.02,0.34536136){\color[rgb]{0,0,0}\makebox(0,0)[lt]{\begin{minipage}{0.2308479\unitlength}\centering Normalization\end{minipage}}}%
    \put(-0.04799387,0.31289975){\color[rgb]{0,0,0}\makebox(0,0)[lt]{\begin{minipage}{0.30423001\unitlength}\centering $\beta_2, \gamma, T_0$\end{minipage}}}%
    \put(0.02456421,0.225){\color[rgb]{0,0,0}\makebox(0,0)[lt]{\begin{minipage}{0.13759147\unitlength}\centering INFT\end{minipage}}}%
    \put(0.22359398,0.36){\color[rgb]{0,0,0}\rotatebox{90}{\makebox(0,0)[lt]{\begin{minipage}{0.07137373\unitlength}\centering \hspace*{.35cm} I/Q\\ modulator\end{minipage}}}}%
    \put(0.01,0.56){\color[rgb]{0,0,0}\makebox(0,0)[lt]{\begin{minipage}{0.29199965\unitlength}\centering 1550 nm\\ $\Delta\nu=1$ kHz\end{minipage}}}%
    \put(0.35369633,0.2657937){\color[rgb]{0,0,0}\makebox(0,0)[lt]{\begin{minipage}{0.32381295\unitlength}\centering $3\times 81.3$ km SSMF\end{minipage}}}%
    \put(0.7406231,0.38){\color[rgb]{0,0,0}\makebox(0,0)[lt]{\begin{minipage}{0.22441998\unitlength}\centering Coherent\\ Receiver\end{minipage}}}%
    \put(0.75,0.50){\color[rgb]{0,0,0}\makebox(0,0)[lt]{\begin{minipage}{0.30\unitlength}\centering 1550 nm\\ $\Delta\nu=1$ kHz\end{minipage}}}%
    \put(-0.02433568,0.45){\color[rgb]{0,0,0}\makebox(0,0)[lt]{\begin{minipage}{0.23390549\unitlength}\centering DAC\\ 88 GSa/s\end{minipage}}}%
    \put(0.285,0.30){\color[rgb]{0,0,0}\makebox(0,0)[lt]{\begin{minipage}{0.22396835\unitlength}\centering OBPF, 50 GHz\end{minipage}}}%
    \put(0.41,0.07){\color[rgb]{0,0,0}\makebox(0,0)[lb]{\smash{EDFA}}}%
    \put(0.41493526,0.46261907){\color[rgb]{0,0,0}\makebox(0,0)[lb]{\smash{EDFA}}}%
    \put(0.58434304,0.035){\color[rgb]{0,0,0}\makebox(0,0)[lt]{\begin{minipage}{0.50755963\unitlength}\centering $\beta_2, \gamma,T_0$\end{minipage}}}%
    \put(0.96,0.025){\color[rgb]{0,0,0}\rotatebox{90}{\makebox(0,0)[ct]{\begin{minipage}{0.045787\unitlength}\centering $\{Q_d(\lambda),Q_c(\omega)\}$\end{minipage}}}}%
     \put(0.11,0.12){\color[rgb]{0,0,0}\makebox(0,0)[lt]{\begin{minipage}{0.13759147\unitlength}\centering $\{Q_d(\lambda_k)\}_{k=1}^4$\end{minipage}}}%
     \put(-0.02,0.115){\color[rgb]{0,0,0}\makebox(0,0)[lt]{\begin{minipage}{0.13759147\unitlength}\centering $Q_c(\omega)$\end{minipage}}}%
  \end{picture}%
\endgroup%
\end{minipage}
\caption{\label{fig:setup}Experimental setup with offline NFT-based detection}
\end{wrapfigure}
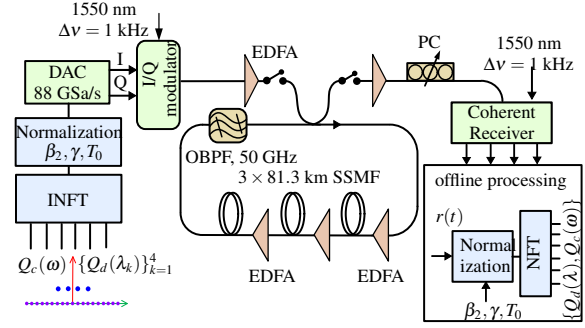
$(1)$ The off-diagonal entries of Eq.~\ref{eq:ZS} can have extremely large variations for $\lambda\in\mathbb{C}^+$ specially when the pulse duration is relatively long. This makes the algorithms too sensitive to the 
over-sampling factor and approximation techniques.\\
$(2)$ Most of the algorithms search for the zeros of $a(\lambda;z)$. It is usually done by some fine 2-D grid search in
$\mathbb{C}^+$ followed by a sub-optimal iterative method like Newton-Raphson zero search method. This requires
lots of $a(\lambda;z)$ evaluations.\\
$(3)$ If the pulse is contaminated by noise, 
the perturbed eigenvalues are correlated. 
Most of NFT algorithms find eigenvalues individually while a joint search is more effective.

The Fourier collocation method is an exception to the above problems, 
but it suffers from a cubic complexity, generating spurious eigenvalues and large numerical errors~\cite{yousefi2014information}.        

\section{The Novel NFT Algorithm: Detection of Eigenvalues from Continuous Spectrum}

\begin{figure*}[tb!]
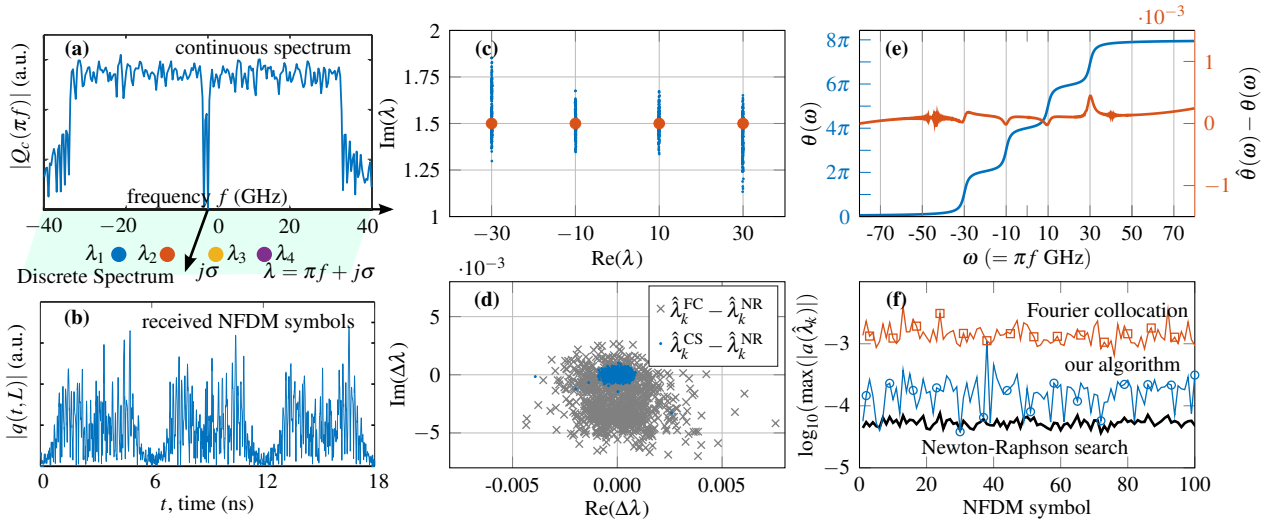

\newlength{\wlength}
\newlength{\hlength}
\setlength{\wlength}{0.3\textwidth}
\setlength{\hlength}{0.15\textwidth}
\definecolor{mycolor1}{rgb}{0.00000,0.44700,0.74100}%
\definecolor{mycolor2}{rgb}{0.85000,0.32500,0.09800}%
\definecolor{mycolor3}{rgb}{0.49412,0.18431,0.55686}%
\definecolor{mycolor4}{rgb}{0.00000,0.49804,0.00000}%
\definecolor{mycolor5}{rgb}{0.50196,0.50196,0.50196}%

\vspace*{-7mm}
\caption{\label{fig:pulse} \textbf{(a)} The nonlinear spectrum of an NFDM pulse \textbf{(b)} The 6 ns received pulses \textbf{(c)} The received eigenvalues, obtained from our algorithm, and the design eigenvalues \textbf{(d)} the gap of estimations of our algorithm (CS) and Fourier collocation (FC) methods to the solutions of Newton-Raphson (NR) exhaustive search. \textbf{(e)} The $\theta(\omega)$
and $\hat{\theta}(\omega)-\theta(\omega)$ after convergence. 
\textbf{(f)} $\max_{1\leq k\leq 4}|a(\hat{\lambda}_k)|$ for 100 NFDM pulses for different eigenvalue estimation methods, our algorithm, FC and NR.  }
\vspace*{-5mm}
\end{figure*}

We present now a new algorithm which retrieves the discrete eigenvalues from the continuous spectrum.
As we will show, our algorithm does not suffer from the above problems.  
Consider an arbitrary pulse with the eigenvalues $\lambda_k=\omega_k+j\sigma_k$, $1\leq k\leq N$
and with the CS $(a(\omega), b(\omega))$ for all $\omega\in\mathbb{R}$. 
It is shown in \cite[P. 49, Eq. 6.27]{faddeev2007hamiltonian} that,
\begin{equation}\label{eq:a_b_relation}
a(\omega) = \mathsf{A}[b(\omega)]\prod_{k=1}^N\frac{\omega-\lambda_k}{\omega-\lambda_k^*},
\text{ with } \mathsf{A}[b(\omega)]=\sqrt{1-|b(\omega)|^2}\exp\left(\frac{j}{2} \mathbb{H}\left[\ln(1-|b(\omega)|^2)\right]\right)
\end{equation}
where $\mathbb{H}[\cdot]$ denotes the Hilbert transform. 
Knowing $b(\omega)$, we can obtain $\mathsf{A}[b(\omega)]$. Then, from $a(\omega)$ and $\mathsf{A}[b(\omega)]$
we can express an all-pass filter $G(\omega)$, i.e. $|G(\omega)|=1$, depending only on the eigenvalues,  
\begin{equation}\label{eq:allpass}
G(\omega)=\frac{a(\omega)}{\mathsf{A}[b(\omega)]} = \prod_{k=1}^N\frac{\omega-\omega_k-j\sigma_k}{\omega-\omega_k+j\sigma_k},
\text{and its unwrapped phase }
\theta(\omega) 
=2\sum_{k=1}^{N} \text{arccot}\left(\frac{\omega_k-\omega}{\sigma_k}\right).
\end{equation}
\textbf{Remark.} An immediate result is on the number of eigenvalues:  $N=\frac{1}{2\pi}\lim_{\omega\to\infty} \theta(\omega)-\theta(-\omega)$ (see Fig.~\ref{fig:pulse}(e)).\vspace{5pt}\\
For a given pulse, we compute $\theta(\omega)$ numerically by
computing $(a(\omega),b(\omega))$ and accordingly, $\mathsf{A}[b(\omega)]$.
The goal is to find $\omega_k$ and $\sigma_k$ from 
the measured phase $\theta(\omega)$ using Eq.~\ref{eq:allpass}. 
Phase synthesis of an $N-$order all pass filter is a 
known problem in signal processing, e.g. \cite{lang1994simple}.
Let $\hat{\lambda}_k=\hat\omega_k+j\hat\sigma_k$ 
denote the estimated eigenvalues for $1\leq k\leq N$.
Define
\begin{equation*}
\hat{\theta}(\omega) 
=2\sum_{k=1}^{N} \text{arccot}\left(\frac{\hat\omega_k-\omega}{\hat\sigma_k}\right)
\text{, and the estimation error }
E(\theta(\omega),\hat{\theta}(\omega))=\int_{-\infty}^{\infty}C(\omega)(\hat{\theta}(\omega)-\theta(\omega))^2 \text{d}\omega,
\end{equation*}
which is a weighted square error with 
some suitable positive weight function $C(\omega)$.
We will have $\hat\lambda_k =\lambda_k$ for $1 \leq k\leq N$, if and only if
the error becomes $E(\theta(\omega),\hat{\theta}(\omega))=0$.
Therefore, we need to find $\hat\lambda_k$ which 
minimizes $E(\theta(\omega),\hat{\theta}(\omega))$. To do this,
one straightforward approach is to apply the iterative gradient descent algorithm,
%
\begin{equation*}
\hat\omega_k^{(m+1)}=\hat\omega_k^{(m)}+\alpha_m \int\frac{C(\omega)\hat\sigma_k^{(m)}}{M(\omega,\hat\omega_k^{(m)},\hat\sigma_k^{(m)})}(\hat{\theta}(\omega)-\theta(\omega)) \text{d}\omega,
\hat\sigma_k^{(m+1)}=\hat\sigma_k^{(m)}-\alpha_m \int\frac{C(\omega)(\omega-\hat\omega_k^{(m)})}{M(\omega,\hat\omega_k^{(m)},\hat\sigma_k^{(m)})}(\hat{\theta}(\omega)-\theta(\omega)) \text{d}\omega,
\end{equation*}
where
$M(\omega,\hat\omega_k,\hat\sigma_k)=\hat\sigma_k^2+(\omega-\hat\omega_k)^2$ and
small positive constants $\alpha_m$. The initial guesses $\hat\omega_k^{(0)}$ and $\hat\sigma_k^{(0)}$ depend on the problem in hand. For instance, we use the design eigenvalues for our pulses in the next section. For simplicity, we also set $C(\omega)=1$ over a large range of $\omega$ though $C(\omega)$
can be further optimized for a faster or more precise convergence.

\section{Performance Evaluation in Experiment}\label{sec:exp}

We verify the performance of our algorithm using the transmission experiment reported before in \cite{aref2018modulation}. Each NFDM pulse is composed of 4 eigenvalues, each modulated 8-PSK independently, and a modulated continuous spectrum by a 64x0.5Gbaud OFDM signal with 32-QAM sub-carriers.
The nonlinear spectrum is visualized in Fig.~\ref{fig:pulse}(a). The 4 eigenvalues are fixed and equal to $\{\pm 3\pi f_0+j\sigma_0,\pm
\pi f_0+j\sigma_0\}$ where $f_0=\frac{10}{\pi}$ GHz and $\sigma_0$ corresponds a fundamental soliton with 
full-width half-maximum (FWHM) of 0.878 ns, 7 times smaller than 
the total duration of 6 ns (with the guard intervals).  
A train of such NFDM symbols is generated 
randomly and transmitted over 18 EDFA amplified spans of 81.3 km standard single-mode fiber. 
The experimental setup is shown in Fig.~\ref{fig:setup}.
The details of the setup, transmission and detection are explained in \cite{aref2018modulation}. 
Here, we focus only on the detection of eigenvalues from the received pulses.

Fig.~\ref{fig:pulse}(e) shows the phase diagram $\theta(\omega)$ in Eq.~\ref{eq:allpass} numerically computed from the continuous spectrum of a received NFDM pulse. We used the trapezoidal NFT algorithm~\cite[Sec. III.A]{aref2016control}
to compute the CS $(a(\omega),b(\omega))$.
We also plot a very small residual error of $\hat{\theta}(\omega)-\theta(\omega)\sim 10^{-4}$ 
after convergence of our least-squared algorithm, showing a precise estimation of zeros and poles of $G(\omega)$.
For most of NFDM pulses, the algorithm required less than 20 iterations.

Fig.~\ref{fig:pulse}(c) illustrates
the estimated eigenvalues $\hat{\lambda}_k$ of 100 received NFDM pulses. We applied also the Fourier collocation (FC) with 512 samples as well as the exhaustive 2-D Newton-Raphson (NR) search with a higher oversampling. Note that these algorithms are much slower than our algorithm. 
To compare the estimation precision of these algorithms, 
we evaluated $|a(\hat{\lambda}_k;z)|$, which must be ideally zero,
using the trapezoidal NFT algorithm with highly over-sampling the pulse.  We observe in Fig.~\ref{fig:pulse}(f) 
that the estimations of our algorithm is more precise than the solutions of FC and slightly less precise than the estimations of NR. Fig.~\ref{fig:pulse}(d) shows that the estimations of our algorithm is much closer to the ones of NR than the estimations of FC. We obtained similar results for different launch powers (the variable power of the continuous spectrum, see \cite{aref2018modulation}).

\vspace*{-1mm}
\section{Conclusion}
We presented a new NFT algorithm for detection of eigenvalues
from the continuous spectrum. The algorithm is fast and more precise than other known algorithm, specially when the pulse has a relatively long duration.

\bibliographystyle{osajnl}
\vspace*{-2mm}
\bibliography{references}

\end{document}